\documentclass[]{mn2e}
\usepackage[dvips]{graphicx}
\usepackage{txfonts}
\usepackage{amssymb}
\usepackage{indentfirst}
\usepackage{epsfig}

\newcommand{\apj}{ApJ}
\newcommand{\apjl}{ApJ}

\newcommand{\aj}{AJ}
\newcommand{\mnras}{MNRAS}
\newcommand{\aap}{A\&A}

\newcommand{\araa}{ARA\&A}

\newcommand{\be}{\begin{equation}}
\newcommand{\bea}{\begin{eqnarray}}
\newcommand{\ee}{\end{equation}}
\newcommand{\eea}{\end{eqnarray}}

\newcommand{\A}{\AA}

\def\OFe{\hbox{O/Fe\ }}
\def\ZFe{\hbox{$Z_{\rm Fe}\ $}}

\def\xmm{{\it XMM-Newton }}
\def\ch{{\it Chandra }}

\begin{document}
\title[X-ray spectroscopy of the hot has in the M31 bulge]{X-ray spectroscopy of the hot gas in the M31 bulge}

\author[Liu et al.]{Jiren Liu$^{1}$\thanks{E-mail: jiren@astro.umass.edu}, Daniel Wang$^{1}$,
Zhiyuan Li$^{2}$, John R. Peterson$^{3}$\\
$^{1}$Department of Astronomy, University of Massachusetts, Amherst, MA 01002\\
$^{2}$Harvard-Smithsonian Center for Astrophysics, Cambridge, MA 02138\\
$^{3}$Department of Physics, Purdue University, West Lafayette, IN 47907
}
\date{}

\maketitle

\begin{abstract}

We present an X-ray spectroscopic study of the nuclear region of the M31 
bulge, based on observations of the \xmm  Reflection Grating 
Spectrometers. The obtained high-resolution grating spectra clearly show
individual emission lines of highly-ionized iron and oxygen, 
which unambiguously confirm
the presence of diffuse hot gas in the bulge, as indicated from previous 
X-ray CCD imaging studies. We model the spectra with detailed Monte-Carlo 
simulations, which provide a robust spectroscopic estimate of the hot gas
temperature $\sim0.29$ keV and the \OFe ratio $\sim0.3$ solar. 
The results indicate that iron ejecta of type Ia supernovae are 
partly-mixed with the hot gas.
The observed spectra show an intensity excess at the OVII triplet,
which most likely represents emission from charge exchanges at the interface
between the hot gas and a known cool gas spiral in the same nuclear region.

\end{abstract}


\section{Introduction}

Early-type galaxies and galactic bulges are mainly made up of old stars and 
evolve passively. Near the end of their life, such stars can collectively 
eject substantial amounts of matter, which is expected to be heated to
the stellar kinematic temperature (typically $\gtrsim 10^6$ K, 
e.g., \nocite{MB03}{Mathews} \& {Brighenti} (2003)). Additional heating by Type Ia SNe can raise the gas 
temperature further.  Ia SNe also provide iron, which can enrich the hot gas
substantially. Such a scenario predicts that the metallicity of
the hot gas should be similar to the stellar metallicity, which is generally
measured to be around solar value \nocite{Tra00}({Trager} {et~al.} 2000).
If the iron ejecta of Ia SNe are well mixed with the hot gas, the resulting iron 
abundance of the hot gas is then expected to be a factor of a few higher. 

However, the expected enrichment of iron from Ia SNe
is not detected in observations of elliptical galaxies. For example, \nocite{Ji09}{Ji} {et~al.} (2009)
analyzed 10 bright elliptical galaxies with both the data from European Photon
Imaging Camera (EPIC) and Reflection Grating
Spectrometers (RGS), and they found near-solar abundances in hot gas.
For some X-ray-faint galaxies, the fitted abundances generally are below
0.1 solar value \nocite{Sar01,Osu04}({Sarazin}, {Irwin} \& {Bregman} 2001; {O'Sullivan} \& {Ponman} 2004).
It seems both the iron ejecta of Ia SNe and the metals in stellar mass loss
are not effectively mixed with, and may be depleted from, hot gas.
But it is so far unclear as to whether such apparent metal discrepancies 
are intrinsic or due to limited quality of the spectral data, 
or due to over-simplified data analysis and modeling.

M31 is an ideal place for a close examination of 
hot gas in a galactic stellar bulge due to the galaxy's
proximity and moderate disk inclination. Prior to \ch and 
{\it XMM-Newton}, however, X-ray observations with very limited sensitivities and
poor spatial resolutions only allowed a crude study
of the X-ray emission in the M31 bulge.
For instance, about 15-26\% of unresolved emission is attributed to faint 
sources based on an extrapolation of the luminosity 
function of detected point 
sources by {\it ROSAT} \nocite{Pri93}({Primini}, {Forman} \& {Jones} 1993), the remaining unresolved emission 
is then {\sl assumed} to arise  from the diffuse hot gas.
Based on a \xmm EPIC MOS1 observation,
\nocite{Shi01}{Shirey} {et~al.} (2001) show a significant soft excess in the unresolved emission
spectrum compared to the averaged spectrum of resolved sources. This 
excess may be due to the diffuse hot gas and/or faint sources with soft spectra. 
Recently, the soft X-ray excess 
in the M31 bulge has been mapped out
with the subtraction of unresolved faint sources by assuming they
spatially follow the stellar K-band light distribution \nocite{Li07,Bog08}({Li} \& {Wang} 2007; {Bogd{\'a}n} \& {Gilfanov} 2008).
The morphology of the remaining diffuse emission is strikingly different from
the stellar distribution and shows a bi-polar shape along the minor axis of the
bulge. This strongly suggests that the diffuse emission represents 
a hot gas outflow from the bulge. To firmly establish the diffuse hot 
gas nature of the soft X-ray emission, however, one needs
to detect the expected line emissions. 

The RGS on-board \xmm have the unique
capability to allow for high-resolution X-ray spectroscopy of moderately 
extended sources because of their large dispersion power. This is the case
for the diffuse hot gas in the M31 bulge. 
To properly interpret the observed spectra, however,
a detailed spatial modeling is necessary, because 
the hot gas is spatially extended and the instrument response depends on
the detector position.
We use detailed Monte-Carlo simulations to model the observed RGS spectra,
which can be decomposed into three components: 
the hot gas, the point sources, and the background. This then allows us to 
characterize the thermal and chemical properties of the hot gas, which 
is the focus of the present paper.

The paper is structured as follows. We describe the observations and our
spectral data modeling method in \S 2. The spectral results are shown in
\S 3. We discuss possible systematic uncertainties in \S 4 and the
implications of our results in \S 5. The errors quoted are at the 90\% confidence
level.

\section{Observation data and modeling method}

\subsection{Observation}

\begin{table}
\caption{\xmm RGS observations \label{tbl:obs}}
\begin{tabular}{cccc}
\hline
ID & t (ks) & $t_{eff}$ (ks) & Obs. date  \\
\hline
0112570401 & 46 & 29 & 2000-06-25 \\
0109270101 & 56 & 21 & 2001-06-29 \\
0112570101 & 64 & 48 & 2002-01-06 \\
\hline
\end{tabular}
\begin{description}
%  \begin{footnotesize}
%\setlength{\itemsep}{-1mm}
\item[] Here t is the exposure time while $t_{eff}$ is the useful time 
after removing intense flare periods.   
%  \end{footnotesize}
\end{description}

\end{table}

We use three archival \xmm RGS observations listed in Table \ref{tbl:obs}.
The total effective exposure time is $\sim$ 100 ks after removing intense flare periods. 
The most recent version of Science Analysis System (SAS, 9.0) is used for
the reduction of photon events.

\begin{figure}
\centerline{\psfig{figure=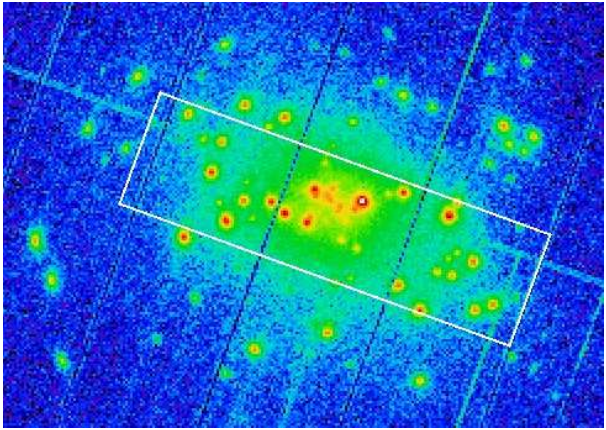, width=8cm}}
\caption{RGS dispersion direction over-plotted on a 0.2-12 keV
EPIC pn image of the M31 bulge. The long axis direction illustrates the dispersion direction
while the short axis is $4'$, which is determined by the field coverage of the
RGS. The plotted box region
is used to extract an EPIC pn spectrum shown in \S 2.2.}
\label{fig:reg}
\end{figure}

Two of the three \xmm telescopes are equipped with the RGS,
each of which is a slit-less dispersive spectrometer. 
Photons from extended sources
are recorded on CCD detectors with the dispersion angle and 1D spatial
information along the cross-dispersion direction. 
The differences of the RGS dispersion directions among the three observations are smaller than
$6.5^{\circ}$, and thus, their spectra can be considered approximately from the same source 
region. The difference in the bore-sights of the three observations 
along dispersion direction is $\sim30''$, which corresponds to 0.06 \A, far less than
the emission line broadening ($\sim$1\A) caused by the spatial extent.
We merge the events of different observations together and assume the nominal center of
the source to be located between the bore-sights.
In Figure \ref{fig:reg} we plot the dispersion direction overlaid
on a 0.2-12 keV EPIC pn image for one observation (ID 0112570101). 

RGS covers an energy range from 5 to 38 \A\ with the largest effective area around
15-20 \A. The spectral resolution is $\sim0.14\theta$\A, depending on the 
angular extent of the observed source, $\theta$, in units of arcmin.
The 1D spatial resolution is around $15''$.
As RGS1 and RGS2 are not identical, separate response files are needed for
each unit. Thus we will analyze the RGS1 and RGS2 spectra separately.
We only consider the first-order photons in extracting
the spectra.

\begin{figure*}
\centerline{\psfig{figure=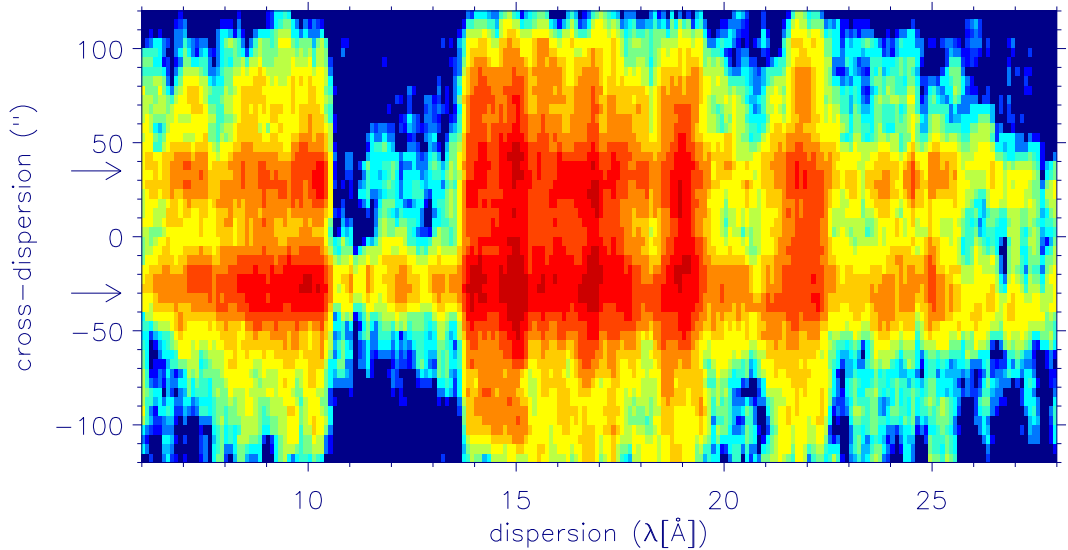, width=15truecm}}
\caption{Cross-dispersion vs dispersion RGS1 CCD image of the M31 bulge.
The trough between 10 and 14 \A\ is due to the failed CCD7 after Sep. 2000.
The positions of two luminous sources are marked with arrows.
Emission lines are clearly seen as vertical contours.}
\label{fig:ccd}
\end{figure*}

Figure \ref{fig:ccd} presents the RGS1 CCD image merged from
all the three observations.
It shows clearly the emission lines of highly ionized iron
(at 15 and 17 \A) and oxygen (at 19 and 22 \A).
The broadening of the lines is mainly due to the spatial extent
of the X-ray emission of the M31 bulge.

\subsection{Modeling method}

For a point source, the intrinsic spectral model can be convolved with a
redistribution matrix function (RMF) and be compared with the observed spectrum
using software like XSPEC (Arnaud 1996). For M31, the observed X-ray photons are from an
extended region, and the RMF depends on the angular position. The 3D
convolution of the angular-dependent RMF is impractical to compute directly, but 
can be evaluated through a Monte Carlo method.
We use the X-ray Monte Carlo code (XMC) developed by \nocite{Pet04}{Peterson}, {Jernigan} \& {Kahn} (2004)
to model the observed RGS spectra of M31. XMC has been used in studies of
elliptical galaxies and galaxy clusters (e.g., Peterson et al. 2001; Xu et al. 2002;
Peterson et al. 2003; Andersson et al. 2009).

Briefly, XMC generates Monte Carlo photons from a spatial plasma model weighted
by its local emissivity, projects them onto the sky, propagates them through an 
instrument, and predicts their detector positions and
energies to be detected by the CCDs. 
The instrument Monte Carlo models the angular-dependent RMF and the details of
its modeling are described in Peterson et al. (2004).
The simulated data are selected in the same way
as applied to the observed data, and the two data sets 
can then be compared with each other.  The process is repeated
by iterating model parameters until an acceptable fit is obtained. In order to reduce the noise
of the Monte Carlo sampling, the number of simulated photons is chosen to be 50 times higher
than the observed photons (29000 for RGS1 and 30000 for RGS2).
The calculation of the gas emissivity calls the XSPEC vapec model
(version 1.3.1, {Smith} {et~al.} 2001).
Since the M31 bulge covers all the field of view of
the RGS detector, we can not extract a background spectrum.
The SAS task rgsbkgmodel over-predicts the background component,
because part of the emission of the M31 bulge
is regarded as the background intensity by rgsbkgmodel.
Thus, we use a semi-empirical model, which includes
soft protons, detector readout noises, and in-flight calibration sources.
The background model is calibrated on blank sky Lockman Hole observations and
is used to randomly generate the background events.
For the full details of the background modeling we refer to Peterson (2003, section 3.4).

The spectrum of the X-ray emission in the M31 bulge has a prominent
power-law component and a soft component
\nocite{Shi01}({Shirey} {et~al.} 2001; Figure 3).
The power-law component comes mostly from %the {\sl Chandra} detected 
low-mass X-ray binaries (LMXB) with X-ray luminosities
$\gtrsim 10^{35} {\rm~ergs~s^{-1}}$,  and a minor fraction ($<10\%$) is
from fainter sources, chiefly 
cataclysmic variables (CV) and coronally active binaries (AB) 
\nocite{Li07,Bog08}({Li} \& {Wang} 2007; {Bogd{\'a}n} \& {Gilfanov} 2008). Such faint sources also contribute
soft photons. But their contribution to the soft excess at energies $<1$ keV
(excluding the power-law component) is estimated
to be about 5\% , based on the low-mass, gas-poor elliptical galaxy of M32.
We neglect this contribution in our analysis 
(more discussion in \S 4.3). Therefore, we
model the observed spectra with three components: 
the diffuse hot gas, the power-law component, and the background.

\begin{figure}
\centerline{\psfig{figure=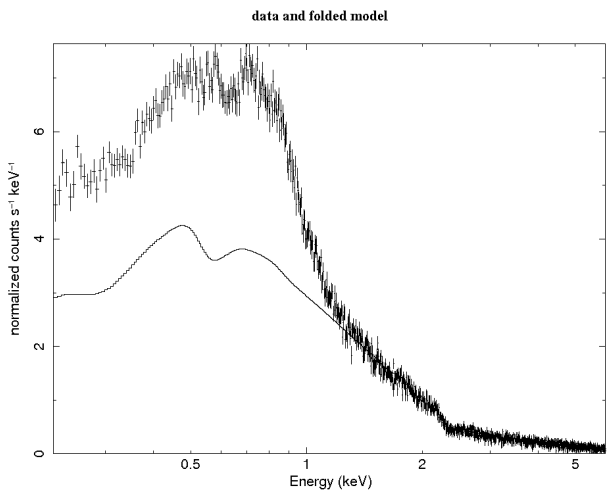, width=8truecm}}
\caption{EPIC pn spectrum of the M31 bulge overlaid with a power-law model.
The spectrum is extracted from the region plotted in Figure 1, and the local
	background is subtracted. The spectrum 
above 1.2 keV is well fitted with a power-law model with index 1.67.}
\label{fig:epic}
\end{figure}

We approximate the electron density distribution with a $\beta$-model
\be
n_e\propto[1+(\frac{r}{r_c})^2]^{-3\beta/2},
\ee
where $\beta=0.49$ and $r_c=54''$ are from the fitting to the surface
density profile of the hot gas \nocite{Li07}({Li} \& {Wang} 2007). 
The normalization of the hot gas is described in the next section.

We characterize the power-law component from an EPIC pn spectrum (Figure
\ref{fig:epic}), which is extracted from 
the observation of ID 0112570101 over the region plotted in Figure 
\ref{fig:reg}. 
The spectrum includes all the emissions from the point sources and the hot
gas. At energies higher than 1.2 keV, the spectrum can be well fitted by a power-low spectrum with index 1.67. This index is assumed for the power law 
component.
As can be seen from Figure \ref{fig:reg}, bright point sources are scattered
around the M31 bulge, and we assume a uniform sphere for the spatial distribution
of the power-law component. The effects of the modeling of the power-law
component are discussed in \S 4.2.
The maximum radii of both the hot gas and the power-law component 
are limited to $8'$, within which the bulk of the X-ray emission arises.
Both components are assumed to be subjected to a foreground absorption of the
Galactic column density
$N_H=6.7\times10^{20}$ cm$^{-2}$ in the direction of M31.

\section{Results}
\begin{figure*}
\psfig{figure=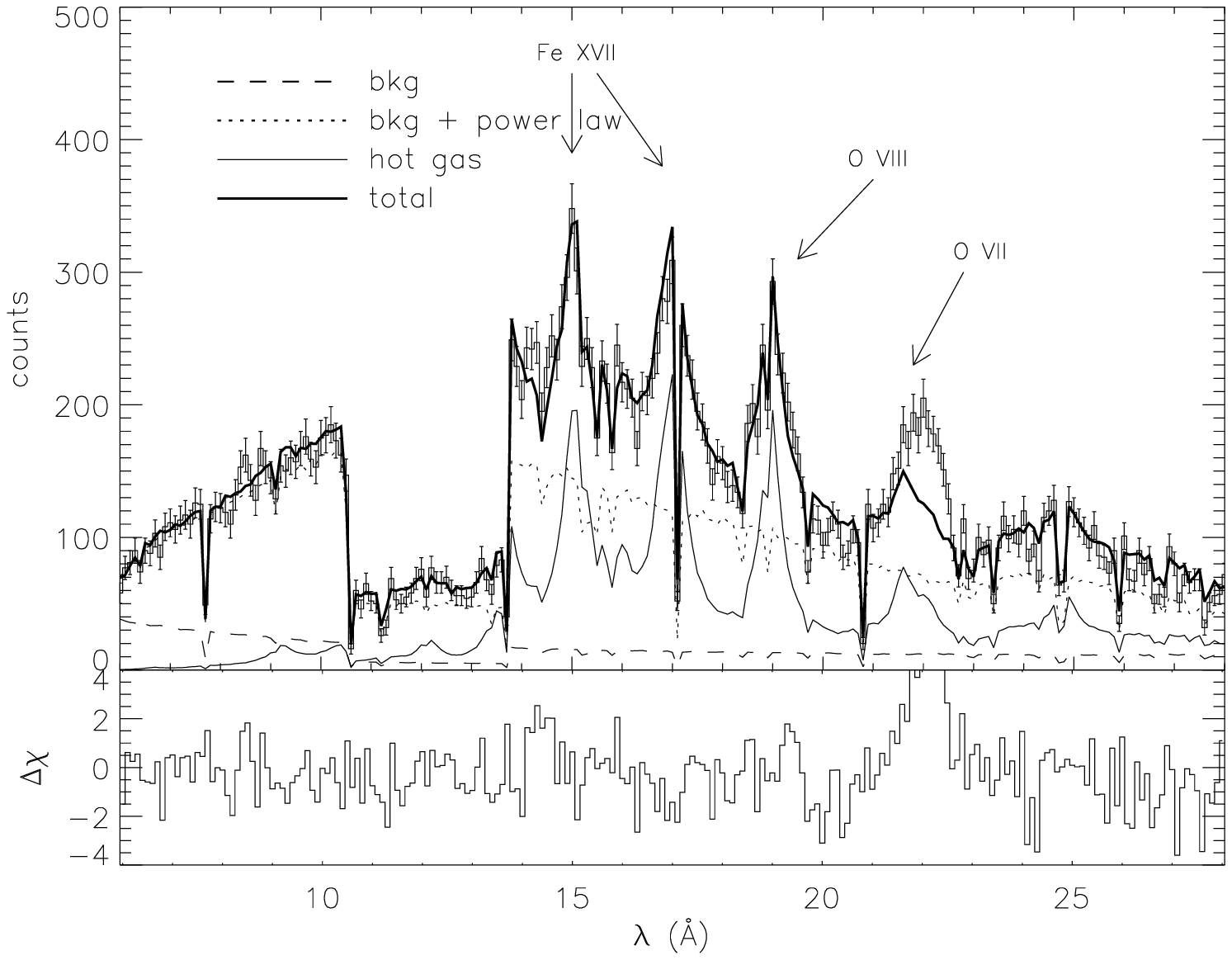, width=8truecm}
\psfig{figure=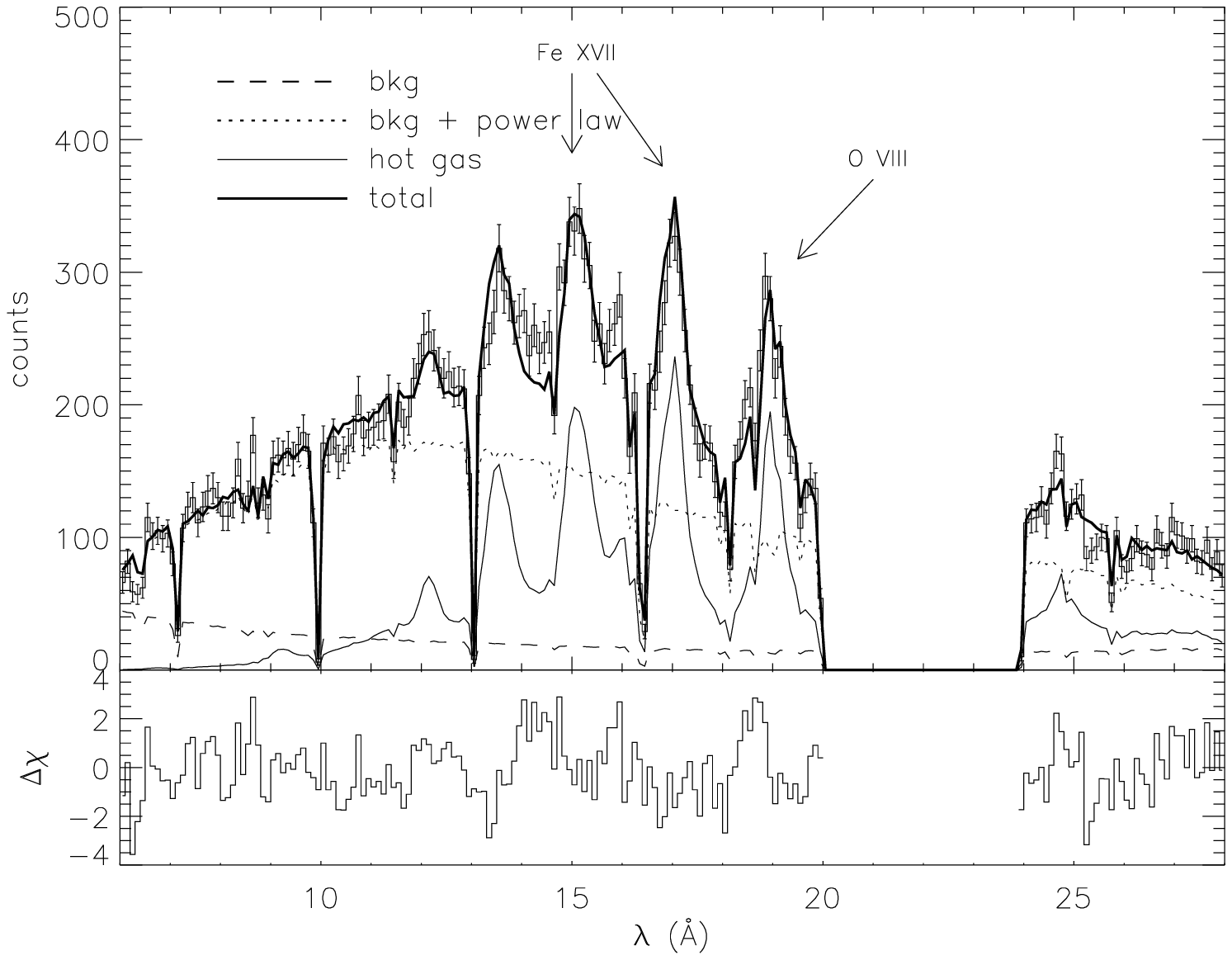, width=8truecm}
\psfig{figure=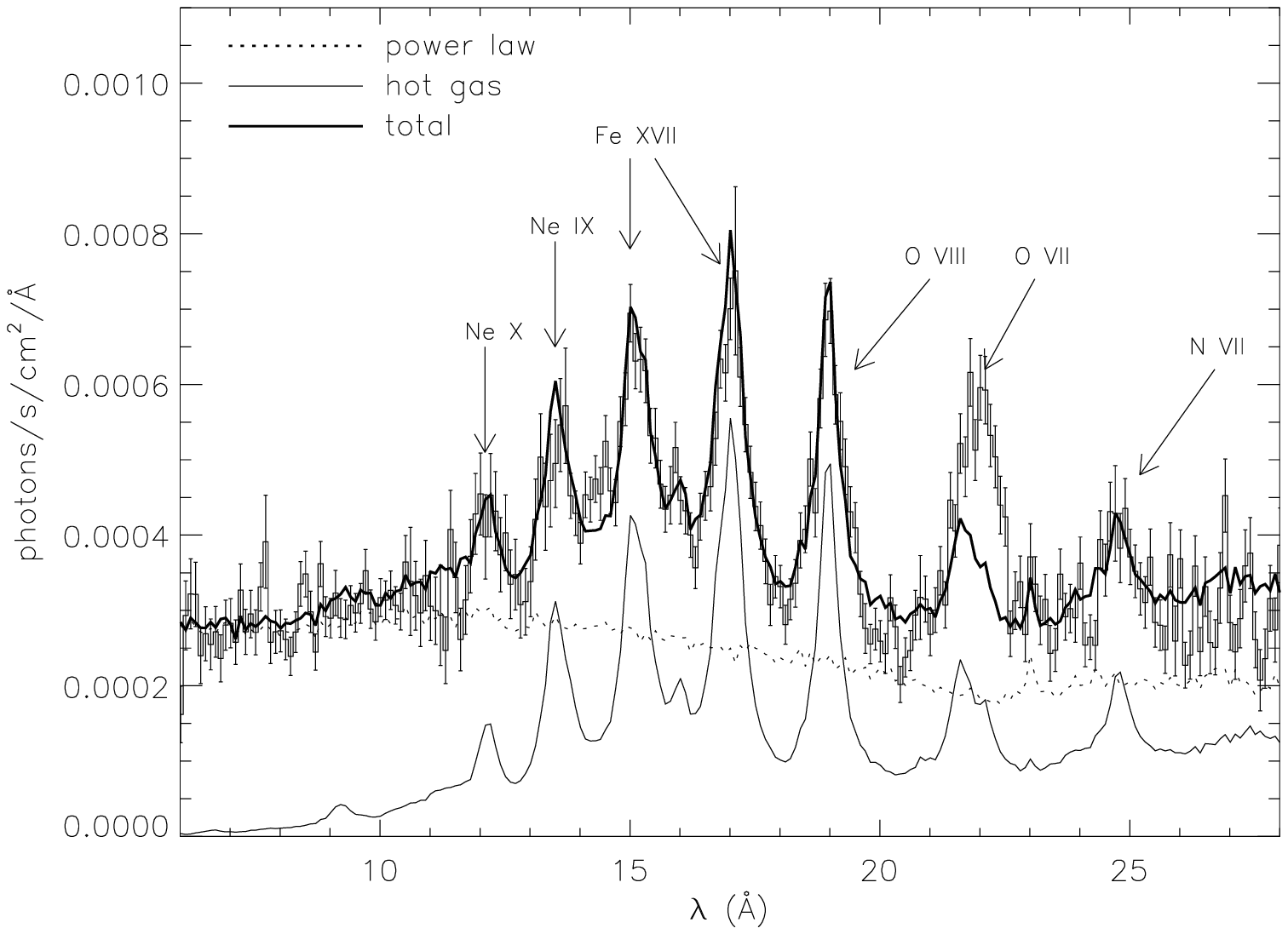, width=8truecm}
%\hspace{1.2truecm}
\psfig{figure=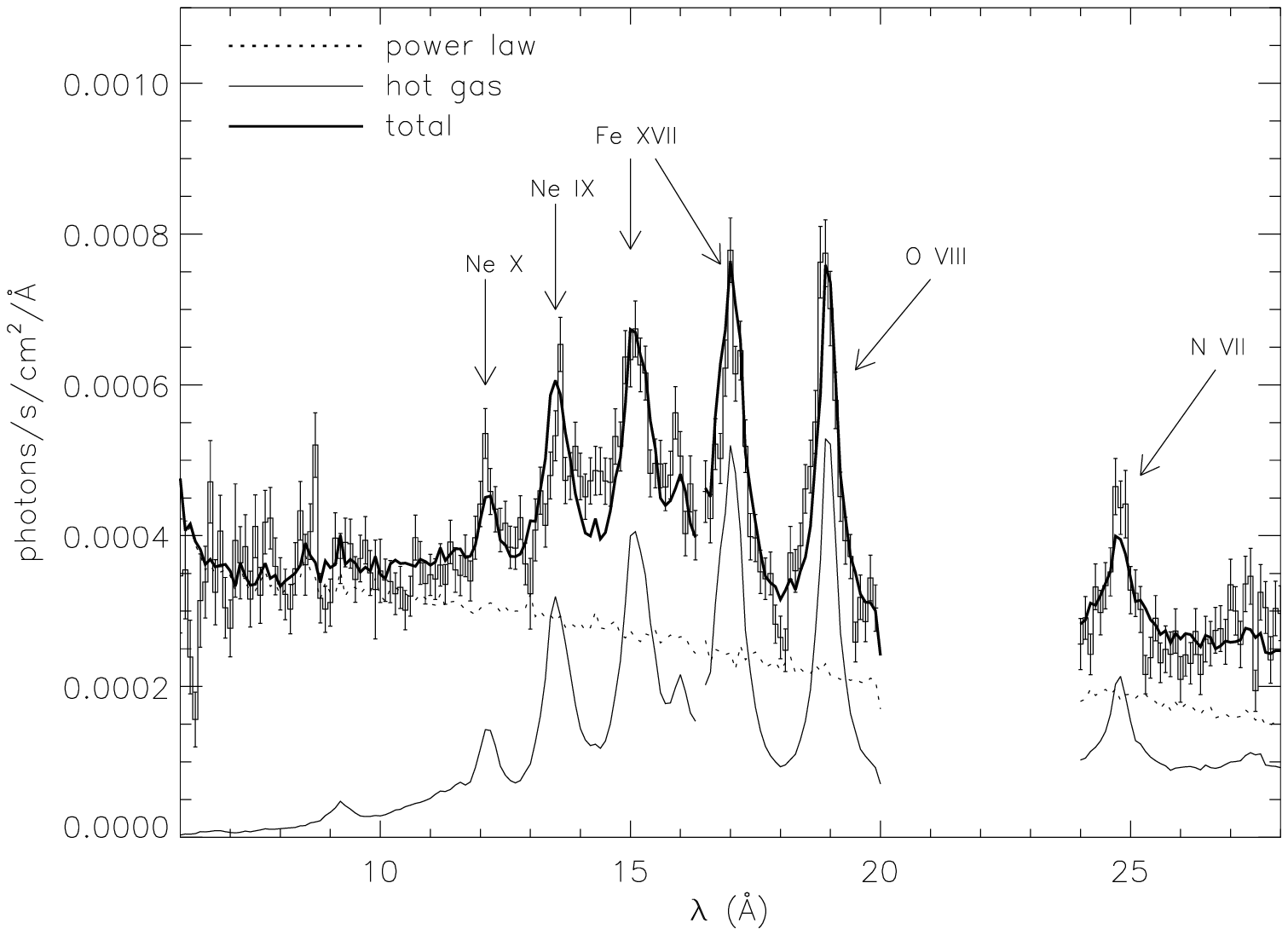, width=8truecm}
\caption{RGS1 (left) and RGS2 (right) spectra of the M31 bulge. 
Top panels are for raw counts, which are used to compare between observed and
modeled data; bottom panels present the background-subtracted spectra corrected for 
the instrument effective area. The histograms are observed spectra.
Note the trough region around 10--14 \A\ of RGS1 is due to the failed CCD7 after
Sep. 2000 and the blank region around 20--24 \A\ of RGS2 is due to the failed
CCD4; the sharp absorption features in the raw counts are due to bad columns 
and gaps between CCDs.} 
\label{fig:spe}
\end{figure*}

Figure \ref{fig:spe} presents the RGS1 and RGS2 spectra of the M31 bulge. 
Note XMC compares the raw counts of the observed and the simulated data in CCD 
detector directly. We concentrate on the wavelength range between 6--28 \A,
which optimizes the signal to noise ratio for our spectral analysis.
Individual emission lines from highly ionized iron and oxygen can be clearly
seen above the continuum, unambiguously confirming the
optically-thin thermal origin for the soft component.

As the spectrum of the soft component is dominated by emission lines 
(Figure \ref{fig:spe}),
the absolute measurement of the metal abundances 
of the thermal plasma is hard to obtain due to their
sensitive dependence on the continuum estimation. In 
contrast, the measurement of an abundance ratio, which is determined by the
ratio of relevant emission
lines, can be more robust. Since the most significant spectral features of the
raw RGS spectra are
from oxygen and iron,
we take \ZFe and \OFe as free parameters. 
The abundances of other elements are assumed to be the same as iron.
(The only other elements showing emission lines are nitrogen (24.8 \A) and
neon (12.1 and 13.7 \A), which will be discussed in \S 4.4.)
Thus, we have a total of five free parameters: two normalization parameters 
$f_p$ and $f_b$, which are
the fraction of total photons from the power-law component and the background, respectively
(the normalization of the hot gas is then $1-f_p-f_b$), and three parameters of the hot gas, T, 
Z$_{\rm Fe}$,
and O/Fe. The abundances are in units of solar values adopted from
\nocite{AG89}{Anders} \& {Grevesse} (1989).
As the photons are from an extended region, the parameters should be
considered as averaged quantities.

We note that the spectra below 10 \A\ are insensitive to the hot gas and 
are mainly determined by the power-law component and the background (Figure
\ref{fig:epic}, \ref{fig:spe}). Since the spectral
shape of both the power-law component and the background are fixed, 
it allows us to separate the normalization
parameters of $f_p$ and $f_b$ from the parameters of the hot gas. 
We first fit $f_p$ and $f_b$ by calculating $\chi^2$ only in the wavelength range of
6--10 \A\ with the parameters of the hot gas listed in Table \ref{tbl:rgs12}. The fit is
insensitive to the exact parameter values of 
the hot gas as long as its contribution to the
spectrum in 6--10 \A\ is small. Then we fix the fitted values of $f_p$ and $f_b$, and fit the
parameters of the hot gas, T, Z$_{\rm Fe}$, and \OFe.
This separation of the parameter fitting slightly underestimates the parameter errors,
but saves computational time substantially, as the Monte Carlo calculation is computationally 
expensive.
The fitting results of both RGS1 and RGS2 data are listed in Table
\ref{tbl:rgs12} and 
plotted in Figure \ref{fig:spe}. The observed RGS1 spectra show excess
emissions around the OVII triplet, which need to be treated separately
(see below), and we thus exclude the region of 21.5-22.5 \A\ in the 
fit to the RGS1 spectrum.

\begin{table*}
\caption{Fitting results}
\label{tbl:rgs12}
\begin{tabular}{c|ccc|cccc}
\hline\hline
 &$f_p$ & $f_b$ & $\chi^2/dof$&T (kev)&\hbox{$Z_{Fe}$}& \hbox{O/Fe}& $\chi^2/dof$\\
\hline
RGS1 &0.57$\pm$0.03& 0.11$\pm$0.02& 45/43 & 0.29$\pm$0.02& 0.13$\pm$0.02& 0.3$\pm$0.03& 236/214 \\
\hline
RGS2 &0.57$\pm$0.03& 0.12$\pm$0.02& 56/43 & 0.28$\pm$0.02& 0.14$\pm$0.02& 0.32$\pm$0.03& 293/185 \\
\hline
\end{tabular}
\end{table*}

Figure \ref{fig:spe}  shows that the RGS1 and RGS2 spectra at energies $\lesssim 10$ \A\  
are well fitted by adjusting the normalization parameters, $f_p$ and $f_b$,
and the emission lines at longer wavelengths also show reasonably good 
fits to the thermal plasma model.
The combination of the three
components generally fits well to the observed spectra except for several
energy regions with large deviations, which we discuss below.
The fitting of the RGS2 spectrum is poorer than RGS1 spectrum.
It is mainly due to the large deviations at 13 and 16 \A, where the CCD
edges locate.

\begin{figure}
\centerline{\psfig{figure=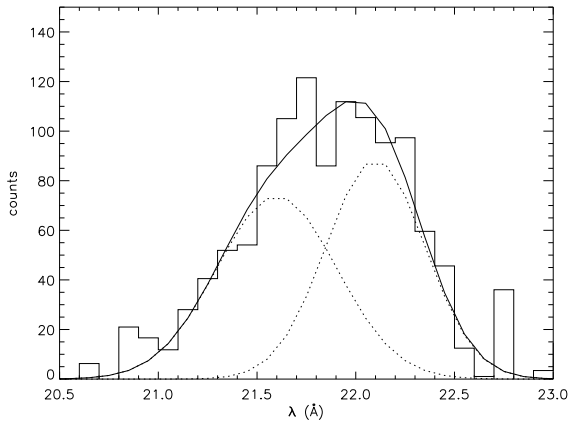, width=8truecm}}
\caption{RGS spectrum of the OVII triplet complex fitted with two Gaussians. 
	The background and the power-law component have been subtracted; a continuum, which
		is calculated as the mean in the line wing regions, has also been subtracted.
	The solid line is the sum of the fitted two Gaussians.}
\label{fig:gau}
\end{figure}

The most apparent deviation of the observed RGS1 spectrum from the modeled one
is the excess around 22 \A, where the helium-like OVII triplet lines occur.
This excess, extending through much of the cross-dispersion range (Figure
\ref{fig:ccd}),
could not be due to any point sources. 
One plausible explanation is that the excess represents
the helium-like OVII triplet lines enhanced by a
charge exchange contribution (see \S 5). 
Figure \ref{fig:gau} presents a decomposition of the observed OVII triplet
complex.  As the forbidden line is blended with the 
resonant line, it is difficult to study the detailed line ratios.
The data also provide no useful constraint on the normally weak
inter-combination line (21.8 \A).
We thus fit the observed line profile with two Gaussians centered on 21.6 and 22.1 \A,
which correspond to the wavelengths of the resonant and the forbidden lines,
respectively.
Both the dispersion and normalization of the two Gaussians are allowed to vary.
The fitted intensity ratio of the forbidden line to the resonant line 
is $1.46\pm0.22$. The ratio of $G=(i+f)/r$ (ratio of the sum of the
inter-combination and the forbidden lines to the resonant line) for a thermal plasma at 
temperature of 0.3 keV is $\sim0.7$ \nocite{Pra82}(e.g., {Pradhan} 1982)
and is $\sim2.2$ for the charge exchange emission \nocite{Bei03}({Beiersdorfer} {et~al.} 2003).
Here the ratio we fitted is approximately $f/(i+r)$, which is smaller than the
ratio $G$. Then the actual ratio $G$ should be larger than 1.46. Thus the 
intensity ratio suggests a multiple-origin of the observed OVII triplet.

\begin{figure}
\centerline{\psfig{figure=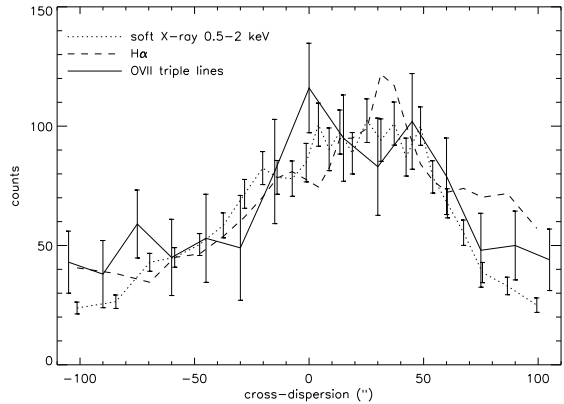, width=8truecm}}
\caption{Intensity profile of the OVII triple 
lines along the cross-dispersion direction. The line intensity is calculated
from the 21-23 \AA\ range after subtracting the continuum estimated from the
intensities detected in the line wing regions in the 20-21 \AA\ and 23-24 \AA\ ranges.
Also plotted are the 1D intensity profiles of the truly diffuse 
emission in the {\sl Chandra} ACIS-I 0.5-2 keV band and the H$\alpha$
emission (Li et al. 2009); both are averaged over a $200''$ region (where the 
H$\alpha$ is significant) along the dispersion direction and scaled to the OVII
intensity at $-60''$. }
\label{fig:ovii}
\end{figure}

Figure~\ref{fig:ovii} compares the 1D intensity profile of the OVII triple 
lines along the cross-dispersion direction with those of the
soft X-ray emission from hot gas and the H$\alpha$
emission from cool gas. The three profiles are 
consistent with each other within the statistical errors, indicating that
these emission components may be closely related.

Another significant deviation of the observed spectrum from the model is near 
14.5 \A. This deviation is possibly due to our simplified modeling of the
power-law component or due to the residual calibration errors, which are
around 10\%. The residual calibration errors are also likely to explain several 
weaker discrepancies between the RGS2 and the model spectrum like the ones near 
13.3 and 16 \A, which occur near CCD edges.

\section{Robustness of the results}

We now examine the robustness of our parameter estimates by considering 
various systematic uncertainties, 
which dominate over statistical errors in the above analysis. The normalization
parameters $f_p$ and $f_b$ depend on the spectral shape of the background and
the power-law component.
Both the temperature and the abundance ratio are mainly determined by the
relative strengths of various emission
lines and are robust.
The absolute iron abundance \ZFe is determined
from the contrast of emission lines to the thermal continuum. Because the emission
lines dominate the thermal spectrum, uncertainties related to the estimation
of the thermal
continuum could affect the iron abundance estimation significantly.
In this section, we discuss the systematic
uncertainties of our analysis and modeling and how they affect the fitted parameters.

\subsection{Uncertainties in the background modeling}

Because the M31 bulge is an extended source and covers all the field of view (FOV) of the
RGS, we can not extract the background spectrum from the same observations.
Thus, we have used a background spectrum based on the deep survey 
observations of the blank sky
of the Lockman Hole. It is possible that this modeled background deviates 
from the real background of the observations used here. 
Nevertheless, the spectral shape of the background is relatively flat and 
not supposed to change with time. Any small deviation should also be absorbed, at least
partially, into the normalization determination of the power-law component.

\subsection{Uncertainties in modeling the power-law component}

The power-law component dominates the spectrum at energies higher than 1 keV
and contributes more than half of the total photons at lower energies. 
A small uncertainty in its modeling will affect the estimation of the
thermal continuum significantly. Many 
point sources scatter around the bulge region and
some of the sources are likely to be time-variable. It is hard to model them
individually for the RGS spectra, and
we have used an averaged spectrum to model their contributions.
Here we study the effect of its modeling by changing the spatial extent and 
the absorption column density.

We take a point source spatial model for the power-law component and re-do the
fitting process. The fitted results for the RGS1 spectrum
are listed in the first row of Table \ref{tbl:point}.
The hot-gas fraction is now 37\%, compared to 32\% for our standard case.
Since the number of the observed photons of emission lines are fixed, 
a higher gas fraction means a larger thermal continuum and hence
a lower abundance: 0.09 solar compared to 0.13 solar. 

\begin{table*}
\caption{Spectral fitting results with various power-law parameters}
\label{tbl:point}
\begin{tabular}{c|ccc|cccc}
\hline \hline
& $f_p$ & $f_b$ & $\chi^2/dof$&T (kev)&\hbox{$Z_{Fe}$}& \hbox{O/Fe}& $\chi^2/dof$\\
\hline
point &0.53$\pm$0.03& 0.1$\pm$0.02& 50/43 & 0.29$\pm$0.02& 0.09$\pm$0.02& 0.3$\pm$0.03& 300/214 \\
\hline
$N_H$=0.04 &0.63$\pm$0.03& 0.11$\pm$0.02& 31/43 & 0.29$\pm$0.02& 0.3$\pm$0.06& 0.31$\pm$0.03& 241/214 \\
\hline
$N_H$=0.1 &0.47$\pm$0.03& 0.12$\pm$0.02& 30/43 & 0.29$\pm$0.02& 0.07$\pm$0.02& 0.31$\pm$0.03& 258/214 \\
\hline
\end{tabular}
\begin{description}
\item Note: $N_H$ is in units of 10$^{22}$ cm$^{-2}$ and point represents the
point source spatial model. 
\end{description}
\end{table*}

Table \ref{tbl:point} also demonstrates the effect due to potential 
uncertainties in the absorption column density $N_H$.
We adopted two values around our adopted Galactic foreground column density
from the HI measurement, which is averaged on a degree scale.
A smaller absorption means a larger contribution of the power-law
component to photons at longer wavelengths and leads to a smaller 
normalization of the 
hot gas. The hot-gas fraction is only 26\% for $N_H=0.04\times10^{22}$
cm$^{-2}$, which results in a high iron abundance of 0.3, while for 
$N_H=0.1\times10^{22}$ cm$^{-2}$, \ZFe is $\sim0.07$.

Clearly, the assumed spatial extent and absorption column density can
significantly change the estimation of the absolute metal abundance.
But both the temperature and the abundance ratio of the hot gas
are hardly affected. This is because the power-law component mainly affects the
continuum estimation, not emission lines.

\subsection{Unresolved soft stellar contribution}

As discussed in \nocite{Li07}{Li} \& {Wang} (2007) and \nocite{Bog08}{Bogd{\'a}n} \& {Gilfanov} (2008), part of the soft X-ray emission should
arise from unresolved CVs and
ABs, the effective temperature of which spans a broad range from 10$^6$ to
10$^8$ K. We estimate their contribution
by scaling the spectrum of the dwarf elliptical M32, which has a 
similar stellar population as the M31 bulge and should contain a 
negligible amount of hot gas
because of the galaxy's low mass. 
The scaling factor is the 
K-band flux ratio of the two galaxies. We find that the 
stellar contribution to the 
soft excess at energies $<1$ keV is only about 5\% for the M31 bulge.
Thus we do not expect significant errors by neglecting this contribution.

\subsection{About other elements}

In previous study, we have tied the abundances of other elements with iron.
As can be seen from Figure 4, the only other two elements showing 
emission lines are nitrogen at 24.8 \A\ and neon at 12.1 and 13.7 \A.
If we allow the nitrogen abundance as free parameter for the RGS1 spectrum, its abundance is 
$0.13\pm0.02$; if we allow the neon abundance as free parameter for the RGS2 spectrum, its
abundance is also $0.13\pm0.02$. That is, the abundances of both nitrogen and neon 
are more close to iron than oxygen. This justifies the grouping method we used
and indicates that oxygen is less effectively mixed with the hot gas than
nitrogen and neon, if they have similar abundances in the stellar ejecta. 
	
\section{Discussion}

We have presented the RGS spectra of the M31 bulge. The observed spectra
show prominent emission lines of highly ionized iron and oxygen and
unambiguously confirm
the thermal origin of the soft excess inferred from previous imaging and
low-resolution spectral analysis. We have
further studied the properties of the hot gas. Our 
measurements of the temperature and the \OFe ratio of the hot
gas are relatively robust,
but the fitted absolute abundance has large uncertainties.

The stellar velocity dispersion of the M31 bulge is 156 km/s 
\nocite{Law83}({Lawrie} 1983), which corresponds to a
kinematic temperature of 0.14 keV. It is only about half of the fitted
temperature of $0.29$ keV. This indicates additional heating,
presumably from Ia SNe.
On the other hand, the main part of the SN energy input is presumably escaped in 
a bipolar outflow as shown in the previous studies 
\nocite{Li07, Bog08}({Li} \& {Wang} 2007; {Bogd{\'a}n} \& {Gilfanov} 2008). 

The effect of the Ia SN feedback is also evident in the abundance pattern.
The spectroscopy of the integrated
stellar light of the M31 bulge shows a super-solar $\alpha$/Fe ratio
\nocite{Puz05}({Puzia}, {Perrett} \& {Bridges} 2005). Theoretical modeling also predicts an enhanced abundance of
oxygen relative to iron for stars in bulges \nocite{Mat99}({Matteucci}, {Romano} \& {Molaro} 1999). 
Thus our fitted \OFe ratio of $\sim0.3$ for the hot gas in the M31 bulge
suggests the iron enrichment  
by Ia SNe. Otherwise, oxygen has to be significantly depleted from the hot gas.

If the iron ejecta of Ia SNe are completely mixed with the hot gas, the expected
iron abundance of the hot gas should then be about 6 times
of the solar value
\nocite{Bog08}({Bogd{\'a}n} \& {Gilfanov} 2008).  While our measurement of \ZFe is not well constrained,
it seems to prefer a sub-solar value.
This is similar to various existing observations of hot gas in faint elliptical galaxies
\nocite{Sar01,Osu04}(e.g., {Sarazin} {et~al.} 2001; {O'Sullivan} \& {Ponman} 2004). 
This iron-discrepancy has invoked the examination of 
the assumption of the complete mixing of iron ejecta \nocite{Fuj97,Bri05}({Fujita}, {Fukumoto} \& {Okoshi} 1997; {Brighenti} \& {Mathews} 2005).
Hydro-dynamical simulations show that the degree of Ia SN ejecta 
mixing with materials from stellar mass loss of evolved stars depends on the
dynamic state of the outflow in the bulge region. In a supersonic outflow case,
for example, the SN ejecta is hardly mixed with the soft X-ray-emitting 
surrounding gases \nocite{Tang09}({Tang} {et~al.} 2009). But in a subsonic outflow case,
the hot and low-density ejecta become increasingly mixed with 
the surrounding medium as they flow outward. This partially-mixing 
scenario helps to explain the unobserved super-solar iron abundance
of the hot gas in the M31 bulge and elliptical galaxies.

However, there are potential other complications. There are 
significant discrepancies in the observational measurements of 
the stellar metallicity of M31 bulge stars. 
While the color-magnitude diagram indicates a metallicity distribution
that shows a peak at the 
solar value with a steep decline at higher metallicities and a gradual
tail to lower metallicities \nocite{Sar05}({Sarajedini} \& {Jablonka} 2005), planetary nebulae appear to have 
a mean oxygen abundance around 0.35 to 0.5 solar value \nocite{Jac99,Sta98}({Jacoby} \& {Ciardullo} 1999; {Stasi{\'n}ska}, {Richer} \&  {McCall} 1998) 
and rarely show super-solar oxygen abundances.
Because planetary nebulae are directly related to the hot gas studied here,
we should probably take their average oxygen abundance of 0.4
solar as the representative value. This, together with our measured \OFe ratio,
results in an iron abundance of $\sim1$ solar for the hot gas. 
This value is still below the expected super-solar iron abundance, supporting
the partly-mixing scenario.

Finally, we have 
presented the evidence for a significant charge exchange contribution
to the OVII triplet. Charge exchange occurs at interfaces between
highly ionized and neutral gases. The presence of neutral
gas is apparent in a spiral structure with an overall dimension of $\sim 4'$.
comparable to the region from which the RGS spectra are extracted.
This structure has been mapped out in mid-infrared dust and H$\alpha$ emissions
and also contains clumps of molecular gas (Li et al. 2009 and references 
therein). The 1D intensity profiles of OVII triplet, the H$\alpha$ emission, 
and the diffuse 0.5-2 keV emission are consistent with each other,
indicating that these emission components may be closely related. 
Similar evidence for charge exchange has also been
presented in the RGS spectrum of the starburst galaxy M82 \nocite{Ran08}({Ranalli} {et~al.} 2008). We are currently
systematically investigating such phenomena and its implications for 
the interplay between the hot and cool gases.

\section*{Acknowledgments}
We thank the referee for useful comments.
This research has made use of \xmm archival data.
XMM-Newton is an ESA science mission with instruments
and contributions directly funded by ESA Member States and the USA (NASA)
This research is partially supported by NASA/SAO through the grant GO8-9088B.

\end{document}